\documentclass[pdflatex,sn-mathphys-num]{sn-jnl}

\usepackage{graphicx}%
\usepackage{subcaption}
\usepackage{multirow}%
\usepackage{amsmath,amssymb,amsfonts}%
\usepackage{amsthm}%
\usepackage{mathrsfs}%
\usepackage[title]{appendix}%
\usepackage{xcolor}%
\usepackage{textcomp}%
\usepackage{manyfoot}%
\usepackage{booktabs}%
\usepackage{algorithm}%
\usepackage{algorithmicx}%
\usepackage{algpseudocode}%
\usepackage{listings}%
\usepackage{tikz}
\usetikzlibrary{arrows.meta,positioning,calc}

\theoremstyle{thmstyleone}%
\theoremstyle{thmstyletwo}%

\theoremstyle{thmstylethree}%

\usepackage{amsmath} 
\usepackage{xspace}  

\newcommand{\cotwo}{\ensuremath{\text{CO}_2}\xspace}
\newcommand{\orfnet}{\ensuremath{\text{oRF}_\text{net}}\xspace}
\newcommand{\orflw}{\ensuremath{\text{oRF}_\text{lw}}\xspace}
\newcommand{\orfsw}{\ensuremath{\text{oRF}_\text{sw}}\xspace}
\newcommand{\netgjkm}{33.7\xspace}
\newcommand{\netci}{20.8, 47.8\xspace}
\newcommand{\olrgjkm}{51.1\xspace}
\newcommand{\olrci}{43.2, 59.1\xspace}
\newcommand{\rsrgjkm}{-17.3\xspace}
\newcommand{\rsrci}{-30.2, -4.1\xspace}

\begin{document}

\title[]{Observational constraints on net radiative forcing confirm aviation contrail warming}


\author*[]{\fnm{Aaron} \sur{Sonabend-W}}

\author[]{\fnm{Scott} \sur{Geraedts}}

\author[]{\fnm{Nita} \sur{Goyal}}

\author[]{\fnm{Joe} \sur{Yue-Hei Ng}}

\author[]{\fnm{Christopher} \sur{Van Arsdale}}

\author[]{\fnm{Kevin} \sur{McCloskey}}

\affil[]{\orgdiv{Google Research}, \orgname{Google}, \orgaddress{
\city{Mountain View}, 
\state{CA}, \country{USA}}}

\abstract{Contrail cirrus represents a critical component of aviation’s non-\cotwo climate impact, but its net radiative forcing, the balance between longwave warming and shortwave cooling, remains poorly constrained by direct observations. As a result, current assessments rely almost exclusively on microphysical models such as CoCiP and global climate simulations. Existing empirical estimates are largely restricted to young, linear tracks, because satellite detection masks have a poor recall of contrails once they spread and merge with natural cirrus, leaving a structural gap in our understanding of long-lived, non-linear contrail cirrus. To address this we use a causal framework that isolates the net radiative contrail effect of flight traffic over the Americas. Building on recent progress that quantified the longwave warming contrail effect using advected flight paths as a proxy for contrails, we expand this continuous treatment approach to capture the highly skewed shortwave cooling impact, delivering a 12-hour lifespan net observational radiative forcing (\orfnet). Our analysis reveals a statistically significant net warming energy forcing of \netgjkm (95\% CI: \netci) GJ/km flown from April 2019 to April 2020, providing a large-scale empirical quantification of long contrail lifespan impact of the same order as, though somewhat larger than, previous bottom-up simulation estimates. This observational benchmark offers an independent line of evidence on the sign and magnitude of the climate impact of contrails.}




\maketitle

\section*{Main}\label{main}

Aviation is one of the fastest growing contributors to climate change~\cite{lee2021contribution}. However, the net radiative impact of contrail cirrus, its largest non-\cotwo climate lever~\cite{lee2021contribution}, remains uncertain due to a lack of empirical constraints, creating a need for independent observational evidence. The net impact of contrails depends on a competing physical balance: they warm the climate by trapping outgoing longwave (LW) radiation, but cool it by reflecting incoming shortwave (SW) solar radiation. Quantifying the net result of these opposing effects is difficult because persistent contrails quickly evolve from distinct linear tracks into spreading, long-lived clouds that are visually indistinguishable from natural cirrus. In this aged phase, they can persist for hours and span thousands of square kilometers, influencing the Earth's energy budget through direct radiative interactions and by altering the environments where natural cirrus would otherwise form.

Consequently, assessing contrail climate impacts has relied almost exclusively on simulation models, ranging from global climate models (GCMs) to specialized microphysical simulations like the Contrail Cirrus Prediction tool (CoCiP) and the Aircraft Plume Chemistry, Emissions, and Microphysics Model (APCEMM)~\cite{schumann2012contrail, fritz2020apcemm}. GCMs attribute a net global effective radiative forcing (ERF) to contrail cirrus of a similar order of magnitude to the aviation sector's \cotwo warming~\cite{lee2021contribution}. However, because the net forcing is a relatively small residual of two large, uncertain opposing fluxes (LW warming and SW cooling) the resulting net climate impact is sensitive to microphysical and environmental uncertainties~\cite{burkhardt2011global}. Microphysics models show that this balance varies by season, latitude, and time of day; for example, nighttime contrails are exclusively warming, whereas daytime contrails induce localized cooling that partially offsets the LW effect~\cite{bock2019contrail}. This microphysical sensitivity is highlighted by recent ensemble evaluations of the CoCiP framework, which show that uncertainty in numerical weather prediction inputs like ambient relative humidity over ice (and internal model parameters) translates into wide spreads of energy forcing estimates~\cite{platt2024effect}. Despite their valuable contributions to understanding contrail microphysics, and guiding avoidance trials \cite{sausen2024, sonabend2024feasibility, Sankar2026, Kirschler2026}, these models lack empirical validation for the aged phase of the contrail lifecycle.

Satellite-based studies generally estimate forcing from three different perspectives. The first one, a ``high-granularity/short-lifespan'' approach, uses computer vision (or manual-annotation) to detect linear contrails and assign satellite pixels that have linear contrail coverage~\cite{mannstein1999operational, minnis2004contrails, ng2023opencontrails}. Relying on these masks restricts the analysis to linear contrails, as detecting aged contrails becomes difficult once they expand and merge with natural clouds. Even when considering linear contrails only, the imperfect precision and recall of contrail detection algorithms introduces noise and possibly systematic bias into contrail climate impact estimates. In this context, establishing a reliable counterfactual is inherently hard. Using pixels adjacent to the contrail is a reasonable choice under the assumption of local smoothness of meteorology~\cite{Navarro2015}, but this can also be challenging: contrails often spread into adjacent natural clouds, making a binary split hard; even in cleanly divided cases, neighboring pixels are not guaranteed to share the same underlying meteorology, surface albedo or other relevant confounders, which can make them poor counterfactual matches.

The second satellite-based framework takes a ``low-granularity/long-lifespan'' approach, bypassing detection challenges entirely through regressions comparing different atmospheric regions. For example, \citet{schumann2013aviation} compared high- and low-traffic regions over decades, while \citet{schumann2021aviation} used the 2020 COVID-19 flight drop as a natural experiment baseline. However, these methods can run into low signal-to-noise ratios because natural weather variability can easily mask the aviation signal within the top-of-atmosphere (TOA) radiative flux data~\cite{wilhelm2021weather}. Additionally, by relying on coarse spatial data, these approaches cannot account for local cloud backgrounds or shifting weather features with high precision, leaving their estimates vulnerable to remaining unmeasured confounding.

\section*{Bypassing Detection Limits via Continuous Air Traffic Tracking}

\begin{figure}[htbp]
    \centering
    \begin{minipage}{\textwidth}
        \centering
        \includegraphics[width=0.95\textwidth]{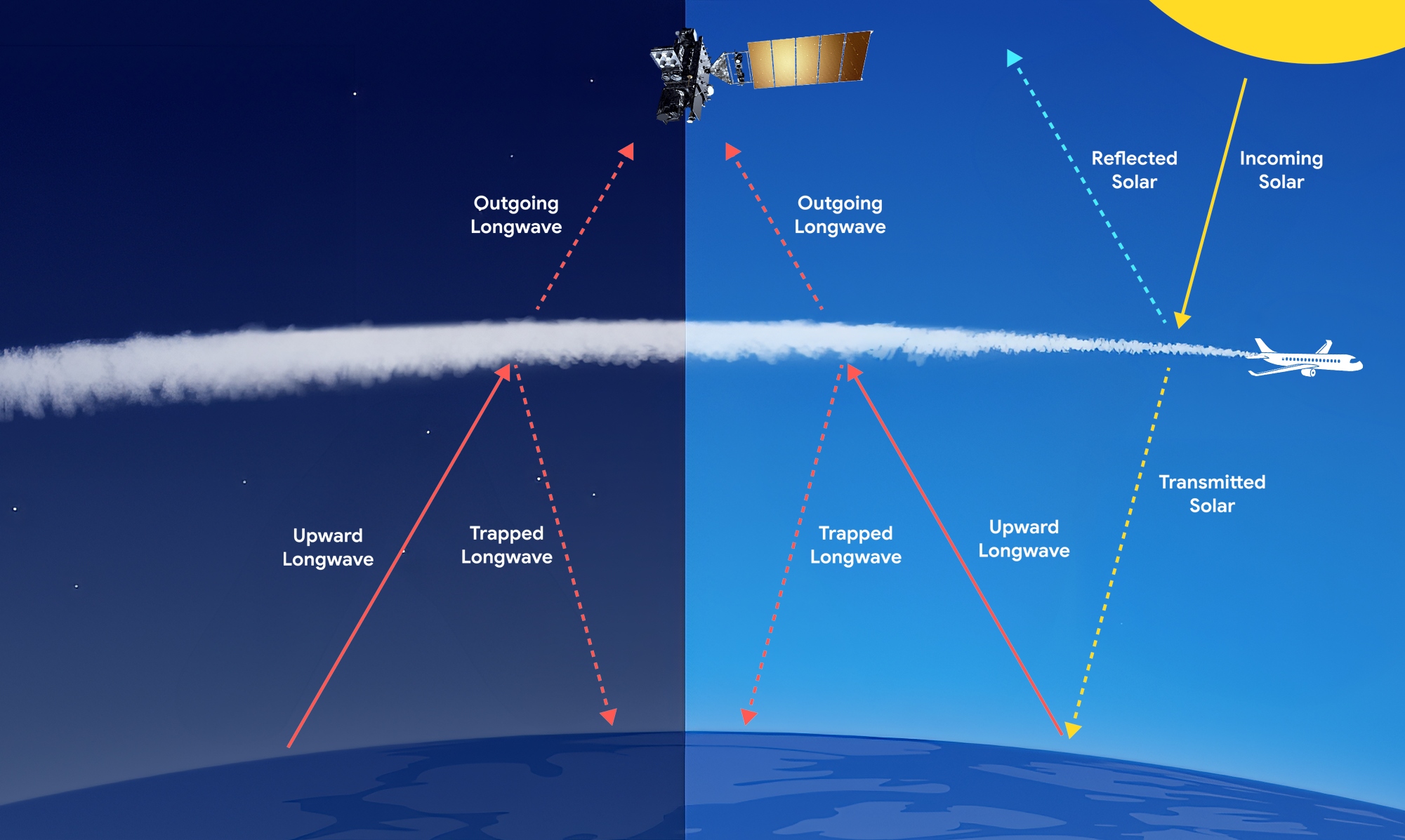}
        \subcaption{Contrails radiative forcing under satellite observation at day and night}
        \label{fig:physical_schematic}
    \end{minipage}
    
    \vspace{12pt} 
    
    \begin{minipage}[c]{0.48\textwidth}
        \centering
        \includegraphics[width=\linewidth]{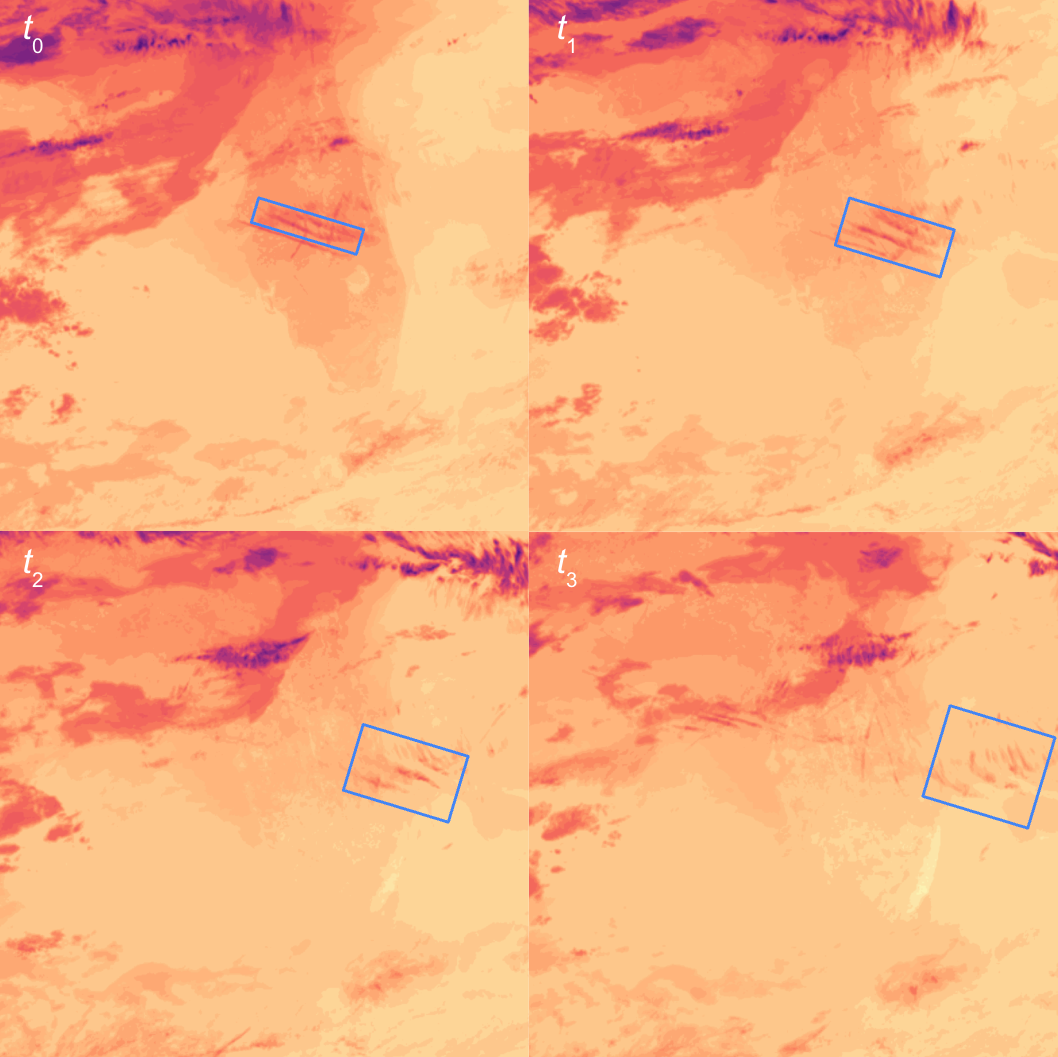}
        \subcaption{COIN OLR over Florida showing a Contrail evolution as it ages, spreads and loses its linear form.}
        \label{fig:coin_contrail}
    \end{minipage}
    \hfill
    \begin{minipage}[c]{0.48\textwidth}
        \centering
        \includegraphics[width=\linewidth]{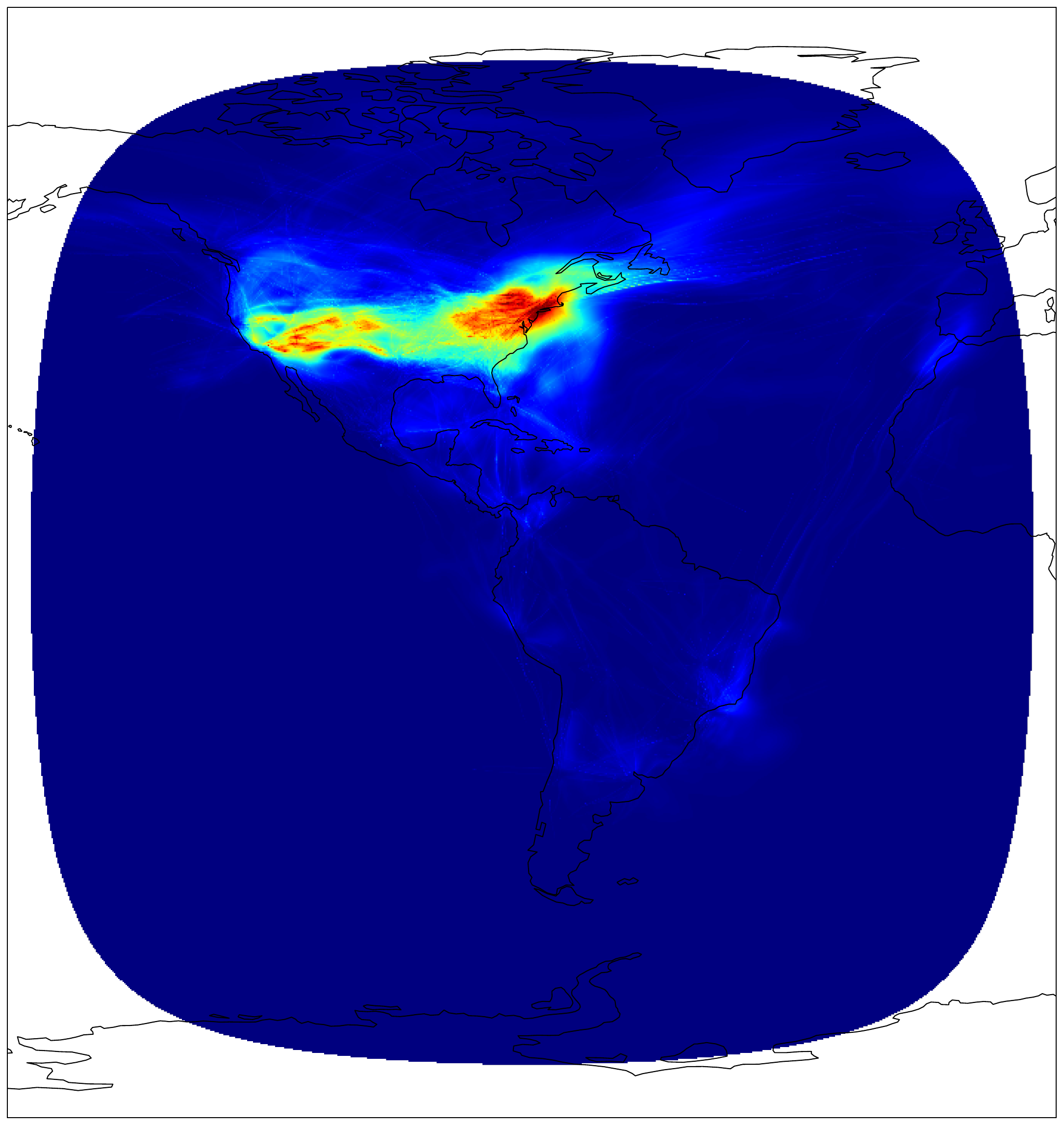}
        \subcaption{Distribution of advected trace density, representing flight kilometers rasterized onto GOES pixels accumulated over a 12-hour window on May 14, 2019, at 01:00 UTC.}
        \label{fig:geographic_domain}
    \end{minipage}
    
    \caption{Traffic-integrated causal framework for isolating contrail radiative forcing. a) Diurnal contrail radiative forcing effects under satellite observation. Nighttime contrails induce an unambiguous net warming effect by trapping upward terrestrial longwave radiation (red arrows). During the day, this longwave trapping is partially balanced by the reflection of incoming shortwave solar radiation (yellow arrow) back to space (blue arrow), driving a variable net radiative response. b) Evolution of a contrail over selected COIN OLR scans over Florida; the contrail can be seen to evolve in time and lose its linear shape while still having a longwave effect. c) Geographic distribution of advected trace density accumulated over 12 hours in the Americas study domain. This spatially continuous exposure variable enables the explicit capture of sublinear radiative saturation from overlapping contrails.}
    
    \label{fig:master_intro_figure}
\end{figure}

In this study, we address these observational limitations by adopting a third ``high-granularity/long-lifespan'' perspective, which shifts from contrail-specific detections towards capturing the radiative forcing effect of flight traffic. Aircraft location data are Automatic Dependent Surveillance-Broadcast (ADS-B) flight path waypoints licensed from FlightAware (\url{flightaware.com}) and Aireon. By mapping this flightpath data directly onto the geostationary satellite view, we preserve high spatial and temporal granularity without aggregating traffic into coarse regional grid boxes. These granular data are rasterized into a continuous pixel-level exposure variable referred to as advected trace density~\cite{sonabend2026observing,Chevallier2023}, which quantifies the accumulated history of flight kilometers per square kilometer of GOES pixel area (km/km$^2$). Using numerical weather model data, these tracks are advected forward over a 12-hour window, encompassing the regions where contrails have traveled and spread across the geostationary satellite view (Fig.~\ref{fig:geographic_domain}). We refer to the Methods section for more detail.

To isolate observational radiative forcing (\orfnet) from natural weather variability and confounding effects, we use a causal inference framework~\cite{sonabend2026observing}. As shown in the directed acyclic graph (Methods Fig.~\ref{fig:causal_dag_panel}), this approach employs a causal regression that uses advected trace density as a continuous treatment. By adjusting for confounders, we can extract the small signal from a dominant background of natural cloud radiative variability.

Modeling radiative effects as a continuous function of advected trace density, rather than restricting the analysis to detected contrails, offers three advantages. First, it comprehensively captures the full set of geostationary-observable contrails contained in the ADS-B data, bypassing the noise and systematic biases introduced when linear detection masks miss faint, aged contrails or falsely classify natural linear-looking clouds. 

Second, it quantifies the marginal change in TOA energy flux per flight kilometer at the pixel level, rather than per detected contrail, and accumulates overlapping traffic within each pixel. Finally, the 12-hour advection window captures rapid atmospheric adjustments, like localized dehydration and the response of natural clouds~\cite{burkhardt2011global}. These adjustments are implicitly observed through satellite imagery but typically omitted in ``offline'' runs of instantaneous radiative forcing simulation models such as CoCiP \cite{schumann2012contrail}.

Leveraging this treatment variable, \citet{sonabend2026observing} use a causal inference model to regress observed OLR as a function of advected trace density, a counterfactual OLR baseline, and background cloud confounders. Adjusting for these background clouds helps clarify the effects of contrails in clear skies versus cloudy scenes. Quantifying the net impact  of contrails requires extending the causal approach to the SW cooling effect. Unlike OLR, Reflected Solar Radiation (RSR) is strictly bounded by incoming solar insolation and is strongly affected by surface albedo, background cloudiness, and solar geometry. We address these physical constraints by modeling COIN RSR with a treatment term that scales the advected trace density effect by the sunlight physically available to be reflected, while explicitly accounting for solar zenith angle and background scene contrast from other confounders.

By combining the previous LW regression with the SW model developed here, we provide a 12-hour integrated net observational radiative forcing (\ensuremath{\text{oRF}_\text{net,H=12}}\xspace) estimate; note for the remainder of this work we will drop the implied ``H=12'' subscript for brevity. Applying this combined approach to the Americas from April 2019 to April 2020, we calculate the distributions of both LW (\orflw) and SW (\orfsw) impacts. The resulting net warming (\orfnet = \orflw + \orfsw) remains substantially (and statistically significantly) non-zero, providing an independent empirical verification of aviation's main non-\cotwo climate impact. 

\section*{Net Observational Forcing of Contrail Cirrus}

\begin{figure}[h]
\centering
\includegraphics[width=0.9\textwidth]{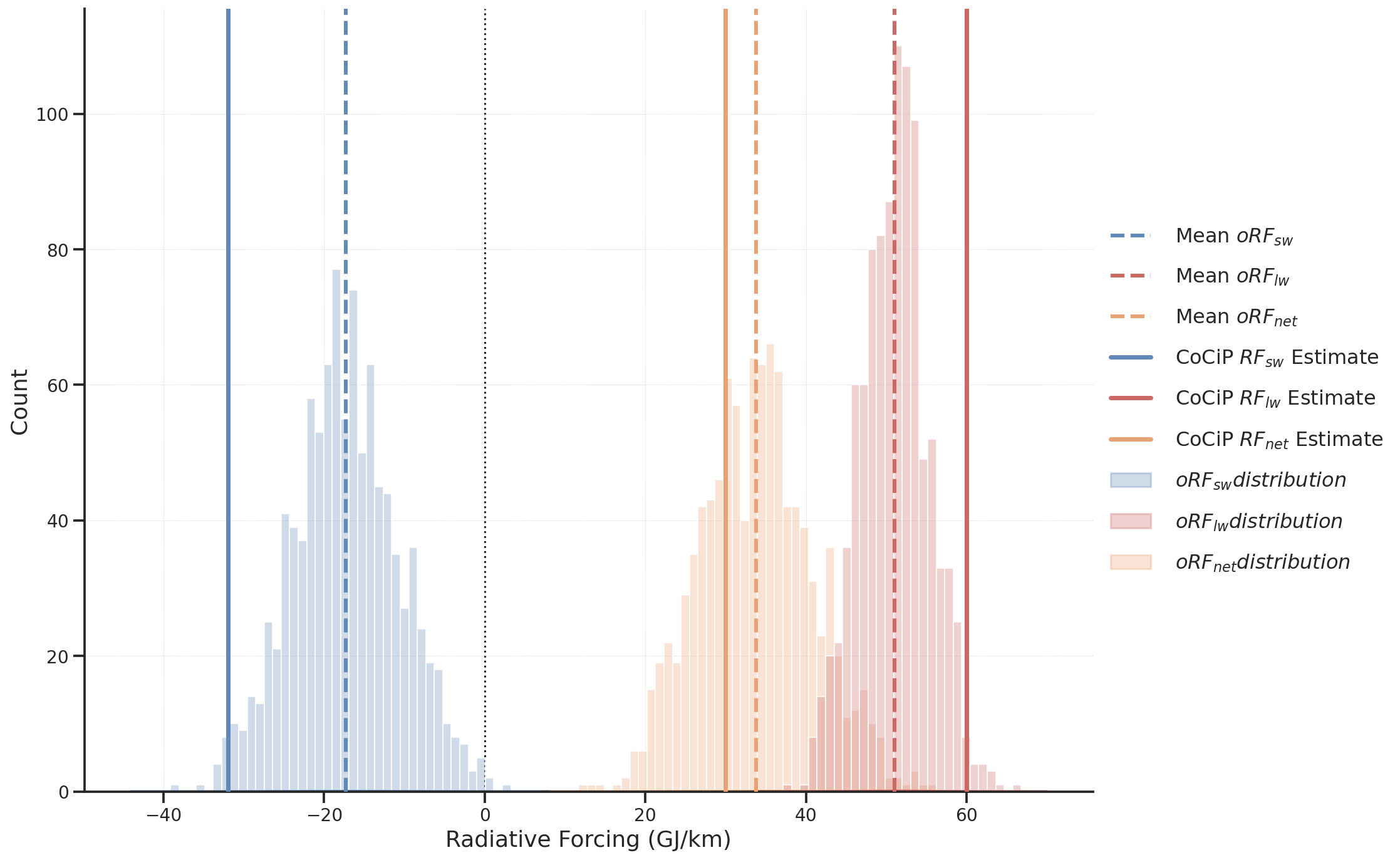}
\caption{Empirical distributions of contrail observed radiative forcing compared with CoCiP estimates. Probability density functions of the 12-hour cumulative observational radiative forcing ($oRF$), derived via block-bootstrap resampling over the Americas domain. Shaded regions represent the distribution of the estimated effects (blue; reflected shortwave radiation, \orfsw), the LW warming component (red; outgoing LW radiation, \orflw), and the resulting net forcing (\orfnet; orange). Vertical dashed lines denote the corresponding theoretical estimates from the Contrail Cirrus Prediction (CoCiP) model for the shortwave (blue), LW (red), and net (orange) components. The solid black vertical reference line indicates zero radiative forcing.}\label{conus_orf}
\end{figure}

We estimate this net observational radiative forcing (\orfnet) over the Americas to be \netgjkm gigajoules per kilometer of flight distance ($\text{GJ km}^{-1}$; $95\%\text{ CI: } \netci$), integrating the total energy forcing effect over the first 12 hours of the contrail lifespan. This net figure is the balance of a LW warming of $\orflw=\olrgjkm\text{ GJ km}^{-1}$ ($95\%\text{ CI: } \olrci$) offset with a SW cooling of $\orfsw=\rsrgjkm\text{ GJ km}^{-1}$ ($95\%\text{ CI: }\rsrci$), with both components estimated from identical resampled days within each of the 1,000 bootstrap replicates, the SW model being fit on the daytime subset of those pixels (Fig.~\ref{conus_orf}). This estimate bypasses the constraints of regional analysis of natural experiments and satellite feature detection of young, linear contrails, and instead establishes a large-scale traffic-based causal constraint for aviation's net climate impact.

When compared against bottom-up microphysical simulations over the same time and
region, our observational estimate is of the same order as the instantaneous
radiative forcing ($\text{iRF}$) estimated by the Contrail Cirrus Prediction
(CoCiP) model, which gives a LW warming of $57.8\text{ GJ km}^{-1}$, a SW
cooling of $-30.2\text{ GJ km}^{-1}$, and a net warming of
$27.52\text{ GJ km}^{-1}$ (Fig.~\ref{conus_orf}). These CoCiP estimates are
computed over the same domain and day set, driven by the same ADS-B flight
inventory and ERA5 meteorology, truncated at the same 12-hour horizon, and
normalized by total flight distance flown rather than by contrail-forming
distance, so that both are expressed per kilometer flown.

Both of our empirical causal-based estimates are smaller in absolute magnitude than their CoCiP counterparts, but not by the same proportion: the LW warming is $12\%$
lower ($\olrgjkm\text{ GJ km}^{-1}$), whereas the SW cooling is $43\%$ weaker
($\rsrgjkm\text{ GJ km}^{-1}$). Because the cooling term contracts almost four
times as much, the residual observational net warming is
$23\%$ larger than the simulated net ($\netgjkm$ against
$27.5\text{ GJ km}^{-1}$), although the CoCiP value still falls within our
bootstrap confidence interval ($\netci$ $\text{GJ km}^{-1}$). The divergence
between the two estimates is therefore concentrated almost entirely in the SW
channel. Part of this difference may reflect the fact that $oRF$ integrates twelve hours of real atmospheric evolution, and so includes the rapid adjustments that
instantaneous forcing calculations omit.

\section*{Isolating Aviation Signals from Natural Variability}\label{isolating_signals}

Applying causal inference to observational data is inherently challenging because there is no ground truth to confirm that the model and its underlying assumptions are correct. To validate their OLR model, \citet{sonabend2026observing} combined observational data with CoCiP simulations to test whether the model could accurately estimate a simulated contrail effect. Specifically, they projected Northern Hemisphere flight traffic onto observed Southern Hemisphere meteorology via an advected trace density latitude-reflection. They then simulated the radiative forcing of these counterfactual contrails and added their LW effect to the observed LW satellite data. This approach created a realistic dataset, demonstrating that the model could successfully recover the simulated contrail forcing when the true effect was known.

\begin{figure}[htbp]
    \centering
    \begin{minipage}[b]{\textwidth}
        \centering
        \includegraphics[width=\linewidth]{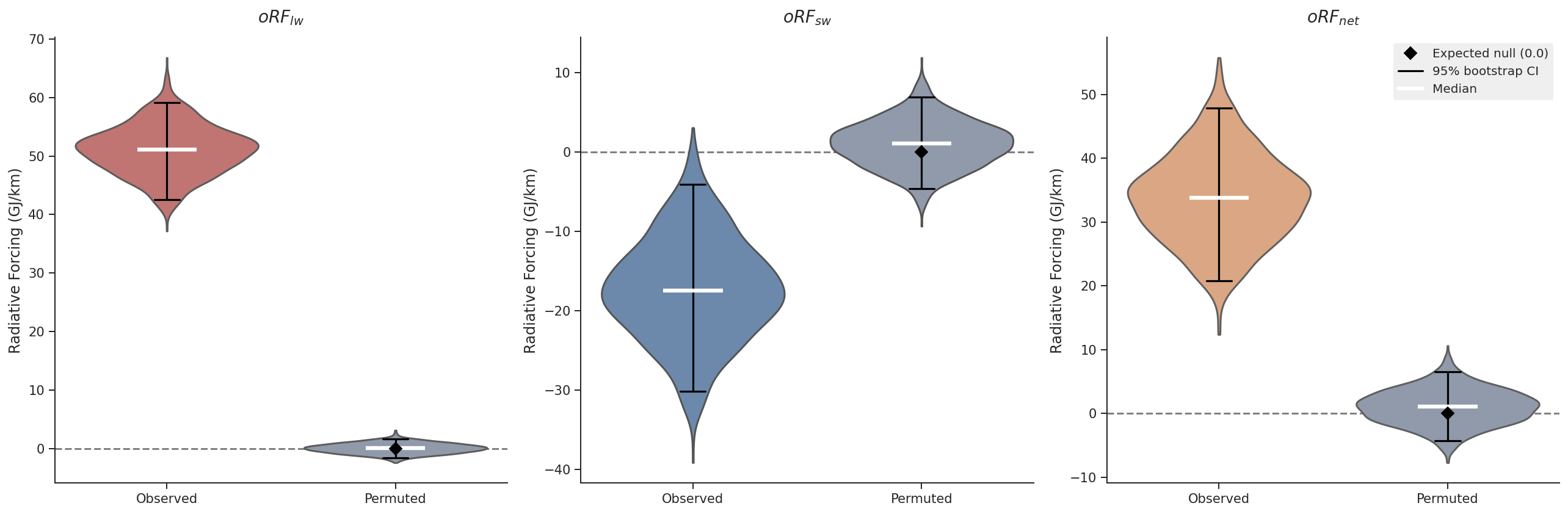}
        \subcaption{Advected Trace Density Permutation Test (Americas Domain)}
        \label{fig:conus_permuted}
    \end{minipage}

    \vspace{12pt} 

    \begin{minipage}[b]{\textwidth}
        \centering
        \includegraphics[width=\linewidth]{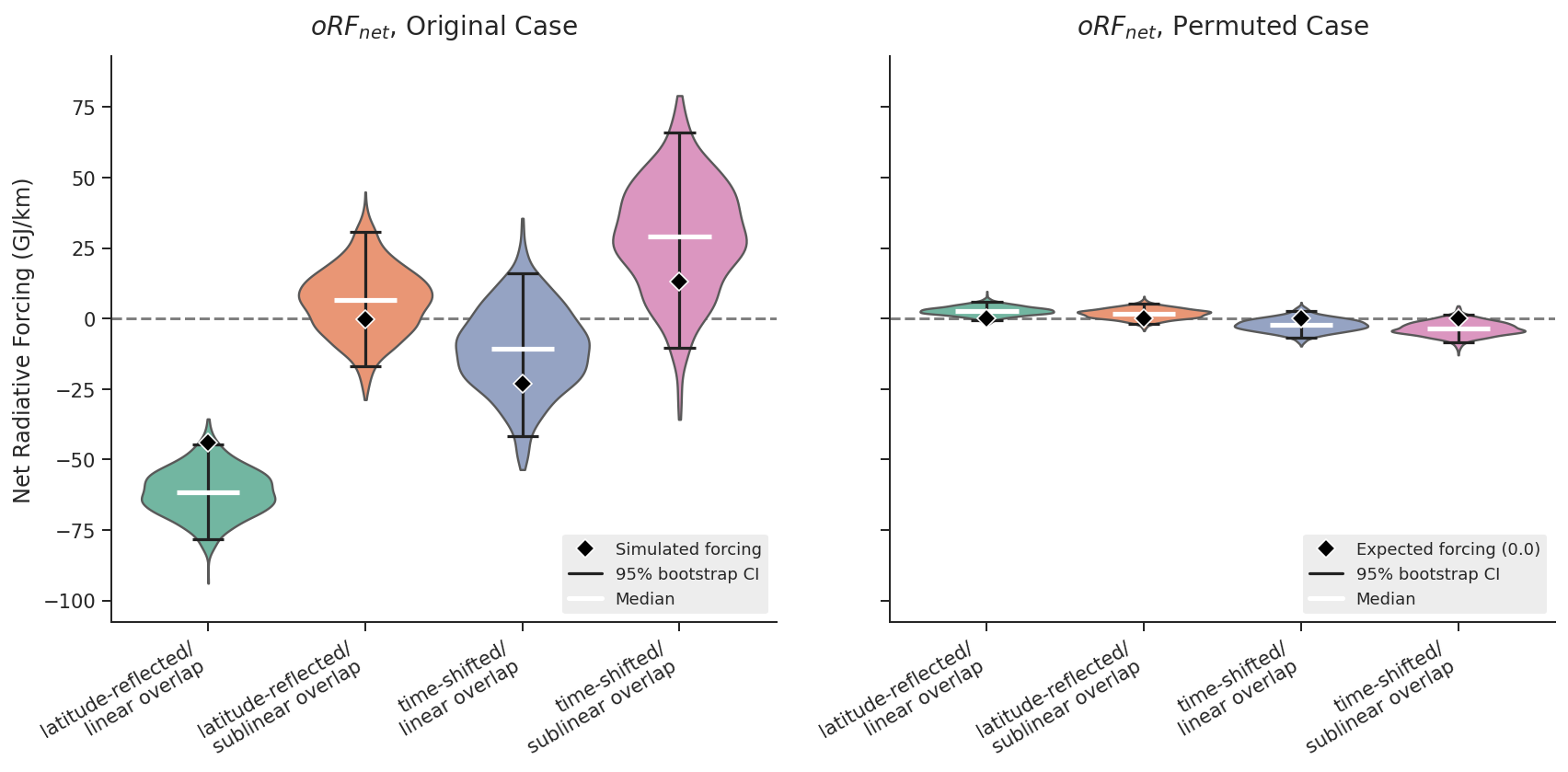}
        \subcaption{Causal Recovery in Synthetic Environments}
        \label{fig:synth_tests}
    \end{minipage}
    
    \caption{Validation of the causal regression framework using synthetic and permuted
test environments. \textbf{(a)} Permutation test on the observed
Americas data: bootstrap distributions of $\orflw$, $\orfsw$ and $\orfnet$ for
the observed advected trace density and for the same field randomly permuted
across the imagery, for which the expected forcing is zero. \textbf{(b)}
Recovery of injected forcing in synthetic environments: bootstrap distributions
of $\orfnet$ for four configurations, defined by how the counterfactual
atmosphere is set up (latitude-reflected or time-shifted) and by how contrail
effects overlap (linear or sublinear); the left panel shows the injected plumes
and the right panel the permuted control. Throughout, black diamonds mark the
target value (the simulated ground truth where forcing is added, and zero
otherwise), horizontal bars the 95\% bootstrap confidence interval, and white
ticks the median.}
    \label{fig:synthetic_validation_matrix}
\end{figure}

The SW effect of contrails is influenced by more confounders than the LW effect. Isolating the causal signal requires a more complex model, which in turn requires a more rigorous validation strategy (see \hyperref[methods]{Methods}). Shifting latitudes inherently breaks the real-world relationships between weather and surface features, so alongside the latitude-reflected environment we developed synthetic testing environments directly within our analysis region that preserve these localized correlations. Specifically, we coupled the observed meteorology with an advected trace density field that was \textit{temporally} shifted forward by four days, while also restricting the advection window to five hours. Both the temporal shift and the five-hour advection truncation break the correlation with real flight traffic, and hence the causal link to real-world contrails, while keeping the observed background surface and weather confounder distributions intact. Any residual correlation with the contrails present in the observed imagery would inflate the recovered effect above the injected forcing; instead, the simulated values fall within the 95\% confidence intervals of the estimates, and permuting the trace density returns an effect indistinguishable from zero (Fig.~\ref{fig:synth_tests}).

Combined, this validation suite spans four distinct synthetic configurations: two background environments (latitude-reflected and time-shifted) crossed with two physical contrail overlap scenarios (linear and sublinear). For each of these four configurations, we evaluate the causal model under two different conditions (Fig.~\ref{fig:synth_tests}). First, in the original case, we test if the model can accurately recover the added simulated forcing. Second, in the permuted case, we randomly shuffle the advected trace density across the satellite imagery against the simulated contrails, removing any true spatial correlation with the radiation fields, to ensure the model does not generate false-positive signals from spurious correlations.

Across all four configurations, when evaluating the permuted cases, the model correctly estimates an effect of near zero GJ/km. The confidence intervals in the permuted cases are narrow in every configuration, an order of magnitude tighter than in the corresponding original cases, so the zero result implies an accurate estimation of the randomized null treatment, rather than a lack of statistical power. Once the advected trace density is de-correlated from the radiance fields, the contrail warming that remains in the scene no longer projects onto the treatment term, and the estimator returns to zero with high precision.

This permuted test is also replicated for the Americas domain on the observed advected trace density used in the real-data causal analysis. Fig.~\ref{fig:conus_permuted} shows that $\orflw$ and $\orfsw$ and thus $\orfnet$ are all successfully estimated to be zero when the treatment is permuted, and that the permuted distributions are narrow in every channel. The permuted intervals are wider in the SW and net channels than in the LW, reflecting the inherently low signal-to-noise ratio of daytime contrail forcing: natural variability in scene albedo and solar reflection generates substantial background noise that is absent at night. These permuted distributions also show the noise floor of the method, and why the SW is a more challenging causal effect to estimate.

Finally, in the original (un-permuted) cases, our model correctly recovers the simulated answer across all forcing scenarios, even in the cases where the net simulated forcing is negative (cooling) as shown in Fig.~\ref{fig:synth_tests}. In all four configurations, the true simulated net forcing is bounded within the 95\% confidence interval of the bootstrapped distribution, and the residual errors do not suggest a systematic bias: the central estimate is biased towards cooling in the latitude-reflected/linear configuration and towards warming in the other three, as expected from sampling noise rather than from a directional flaw in the estimator. Intervals are wider here than in our observational estimate because the synthetic runs are fit on a smaller day sample (36 and 180 days for the latitude-reflected and time-shifted respectively). This consistent performance demonstrates that we reliably isolate the cloud-mediated radiative footprint of aviation, regardless of the sign and magnitude of the forcing, or whether the simulated lifespan was 5 or 12 hours.

\section*{Discussion}

While using advected trace density helps to eliminate the biases of linear detections and regional analyses, our estimates remain bounded by the specific spatial and temporal scope of the observational data. Focused over the Americas, this regional analysis provides a critical empirical anchor but may not fully represent the distinct synoptic meteorology of other major aviation corridors, such as Europe or Asia. Additionally, our 12-hour advection window captures a large portion of the aged contrail lifecycle, but truncates the forcing of long-lived contrails that persist beyond this timeframe. Finally, while the causal regressions adjust for pixel-level confounding, observational data at this scale is fundamentally non-experimental; our estimates rely on necessary causal assumptions to isolate the highly correlated advected trace density from background meteorological variability.

The framework presented here provides an observational constraint on contrail
climate impacts. By establishing a robust statistical baseline to evaluate the aviation industry’s contrail radiative footprint, this approach complements and
constrains existing microphysical models, reducing reliance on purely simulation-based
estimates.

\section*{Methods}\label{methods}

\subsection*{Data Sources}

All features used for the causal regressions are collocated onto the GOES-16 ABI pixel grid for the Americas domain over April 2019 to April 2020. Top-of-atmosphere upwelling LW and SW broadband
fluxes are the COIN estimates derived from GOES-16 ABI narrowband radiances~\cite{MCCLOSKEY2023113376}. Cloud phase is the GOES-16 ABI Level~2 Cloud Top Phase
full-disk product (ACTP)~\cite{pavolonis2010actp}, the categories we use are
clear sky, liquid water, supercooled liquid water, mixed phase and ice; pixels flagged as
unknown phase are dropped from the analysis. Flight paths defined as ADS-B waypoints are obtained from FlightAware
(\url{flightaware.com}) and Aireon.

Meteorological covariates are obtained from the ERA5
reanalysis~\cite{hersbach2020era5}, read from the analysis-ready ARCO-ERA5 archive at
$0.25^\circ$ hourly resolution~\cite{carver2023arco} and interpolated onto the ABI grid. The
counterfactual LW baseline $\text{OLR}_{\text{ERA5}}$ is top net thermal radiation
(\texttt{ttr}); the SW baseline $\text{RSR}_{\text{ERA5}}$ is the difference between
top incident and top net solar radiation (\texttt{tisr}$-$\texttt{tsr}); the incident
shortwave flux $S$ is \texttt{tisr}, 2 meter surface temperature $T_{2\text{m}}$ is
\texttt{t2m}. The $u$, $v$ and $w$ wind components used to advect flight waypoints are taken from
the same archive at the pressure level matching each waypoint's altitude.

Surface albedo $\alpha_{\text{sfc}}$ is the MODIS white-sky shortwave broadband albedo from
the MCD43C3 version~061 daily climate modeling grid product at $0.05^\circ$
resolution~\cite{schaaf2021mcd43c3}, reprojected onto the GOES-16 viewing perspective. The
MCD43C3 retrieval is defined over land only, so pixels identified as water by the
accompanying land/sea mask are assigned a constant albedo of $0.08$, representative of the
broadband shortwave reflectance of open ocean~\cite{jin2004ocean}. Approximately $30\%$ of the pixels in the Americas
domain are ocean and receive this constant. The solar zenith angle $\theta_0$ is computed
analytically from each pixel's latitude, longitude and observation local time.

\subsection*{Observational Data and Counterfactual Baselines}

The empirical framework models top-of-atmosphere (TOA) longwave (LW) and shortwave (SW) radiative fluxes as independent outcomes with separate causal regression models. Both use advected trace density as the continuous treatment variable to estimate the average treatment forcing effects. The SW and LW measured outcome fields are obtained from the Collocated Irradiance Network (COIN)~\cite{MCCLOSKEY2023113376}. This is a neural network model that estimates pixel-wise broadband upwelling fluxes (in $W/m^2$) directly from GOES-16 Advanced Baseline Imager (ABI) narrowband radiances. Operating from a geostationary orbit at $75^\circ\text{W}$ longitude with a 10-minute temporal refresh rate, the ABI provides full-disk imagery consisting of $5424 \times 5424$ pixels, at a nadir spatial resolution of $\sim 2\text{ km}$~\cite{heidinger2020abi}. We refer to this as the “ABI pixel grid”. See Fig. \ref{fig:geographic_domain} for the GOES-16 ABI pixel grid used.

The main counterfactual baseline we use is the TOA OLR and RSR fields simulated by the ERA5 reanalysis~\cite{hersbach2020era5}, following~\citet{sonabend2026observing}, and analogous causal frameworks developed to evaluate aerosol–cloud interactions~\cite{chen2022machine,jia2024revisiting}. These reanalysis fields act as exogenous variables that account for the main meteorological drivers of flux. ERA5 contains no contrail parameterisation, assimilates no contrail-resolving observations, and contrail-scale perturbations are far below the assimilation system's effective resolution~\cite{hersbach2020era5}. Therefore it provides an estimate of the TOA fluxes that would have been observed had there not existed any contrails. To adjust for residual confounding and isolate the advected trace density treatment effect from this baseline variability, additional observation-based confounders are added into each regression model (Fig.~\ref{fig:methodology_combined}). 

Permuting advected trace density across the imagery returns an effect
indistinguishable from zero (Fig.~\ref{fig:conus_permuted}), which shows that
spatial biases in the ERA5 baseline do not confound the treatment effect through
unstructured spatial correlation. Because permutation also destroys the
geographical distribution of air traffic, it cannot by itself exclude a bias that is
systematically aligned with flight traffic patterns. The time-shifted synthetic
environment addresses this case directly: shifting the trace density forward by
four days preserves the geographic structure, since traffic patterns repeat closely
from day to day, while breaking the link to the contrails actually present in the
imagery. A geographically-aligned baseline bias would therefore still overlap the shifted treatment and artificially inflate the estimated effect above the simulated forcing. The observed deviations are small relative to the bootstrap uncertainty and do not have a consistent sign across the four synthetic configurations, which provides no evidence of systematic leakage of real-contrail signal into the shifted
treatment.

\subsection*{Advected Trace Density as a Continuous Treatment}

Following the approach used by \citet{sonabend2026observing}, flight trajectories are advected forward for a duration of 12 hours. We scale advection uncertainty with time. The 12 hour truncation implies a bound on the lifespan considered for the net climate impact; radiative forcing may in fact persist beyond 12 hours.

The continuous treatment variable, advected trace density ($A$), is constructed by advecting the flight trajectories, mapping flight kilometers to represent potential contrail kilometers~\cite{sonabend2026observing,Chevallier2023}. These continuous trajectories are broken down into discrete waypoints with three-dimensional coordinates (latitude, longitude, and altitude) along with their corresponding timestamps. Every waypoint is then moved forward step-by-step over a rolling 12-hour window using the $u$, $v$ and $w$ wind components from the ERA5 reanalysis, interpolated at the flight’s altitude and position. This advection is calculated using a three-dimensional Runge-Kutta method, with an additional downward ice crystal sedimentation modeled based on average terminal velocity; for further details see~\cite{geraedts2024scalable}.

To account for errors in the wind fields and location uncertainty, each advected waypoint is assigned a Gaussian density centered on its expected location. The standard deviation of this distribution expands over advection time at a rate of 12.4 kilometers per hour to reflect the growing uncertainty cone~\cite{sonabend2026observing}. These expanding footprints are projected onto the GOES-16 ABI pixel grid and normalized to represent flight kilometers rasterized within the pixel area, yielding advected trace density units of km/km$^2$. For any given 10-minute satellite frame, the total exposure at a specific pixel is calculated by accumulating all active flight segments that have blown into that pixel's boundaries over the past 12 hours. When multiple flight tracks overlap within the same grid cell, their individual densities are added together, providing a continuous measurement of potential contrail exposure per pixel.

\subsection*{Causal Regression Framework for Longwave and Shortwave Radiative Forcing}

The LW estimator used is from \citet{sonabend2026observing}; we include it
here for completeness. Observed COIN OLR is regressed on advected trace density, the
ERA5 counterfactual baseline $\text{OLR}_{\text{ERA5}}$, and background cloud
phase:

\begin{equation}\label{olr_model}
\mathbb{E}[\text{OLR} \mid A, \mathbf{X}] = \beta^{\text{lw}}_0
  + \beta^{\text{lw}}_1 \text{OLR}_{\text{ERA5}}
  + \sum_{j=1}^{4} \beta^{\text{lw}}_{\text{CP}j} I_{\{\text{CP}=j\}}
  + \sum_{j=0}^{4} \delta_j\, I_{\{\text{CP}=j\}} \cdot A .
\end{equation}

where $I_{\{\text{CP}=j\}}$ are the cloud-phase indicators defined above and
clear sky ($\text{CP}=0$) is the omitted reference category. The treatment
coefficients $\delta_j$ give the marginal change in outgoing longwave flux per
unit of advected trace density within each cloud-phase category. Equation~\ref{olr_model} is fitted ordinary least squares. We refer
to~\citet{sonabend2026observing} for the derivation and for the validation of
this estimator against simulated forcing.

\begin{figure}[htbp]
    \centering
    
    \begin{subfigure}[b]{0.48\textwidth}
        \centering
        \includegraphics[width=\textwidth]{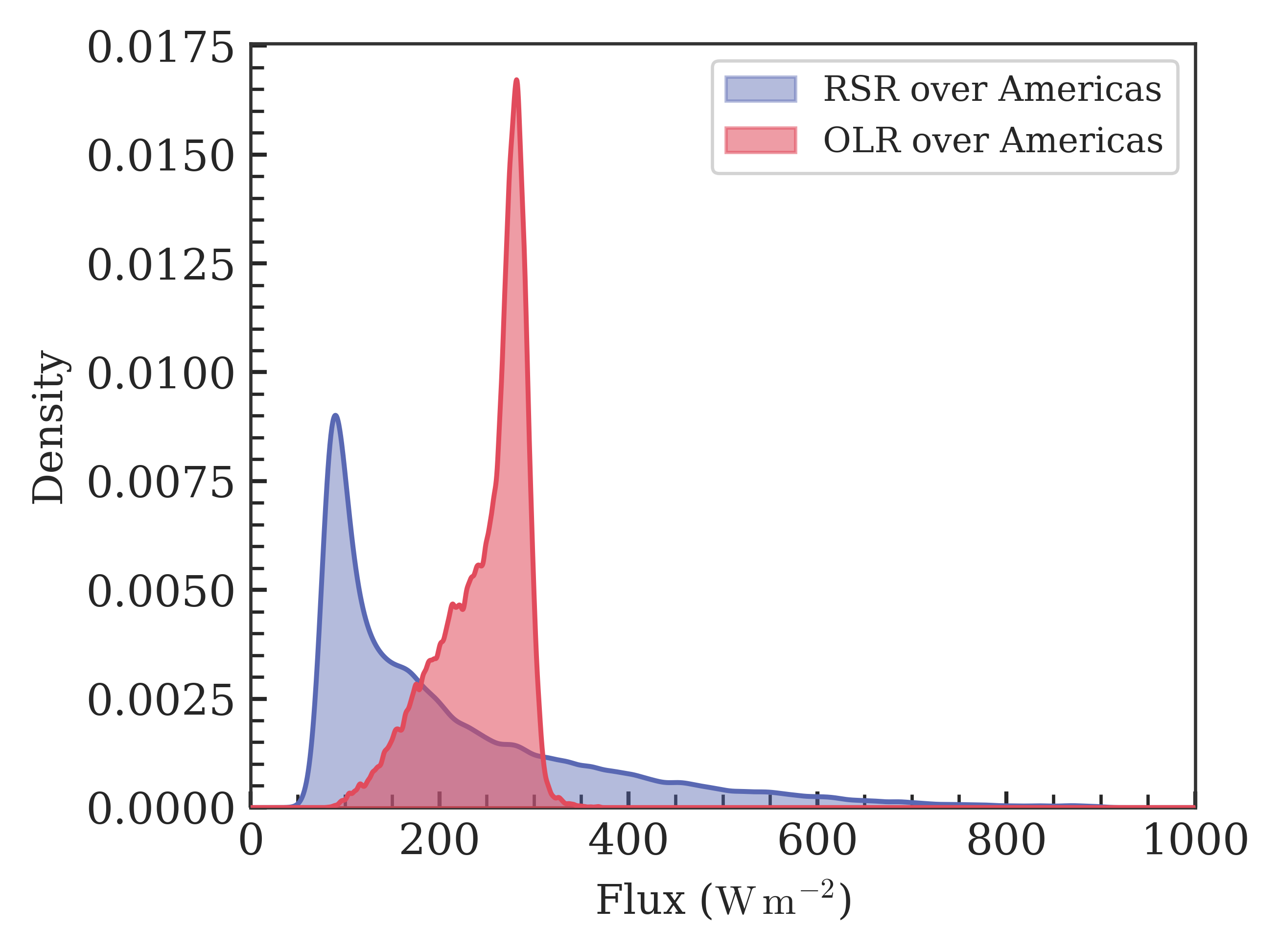}
        \caption{}
        \label{fig:flux_over_conus}
    \end{subfigure}
    \hfill
    \begin{subfigure}[b]{0.48\textwidth}
        \centering
        \includegraphics[width=\textwidth]{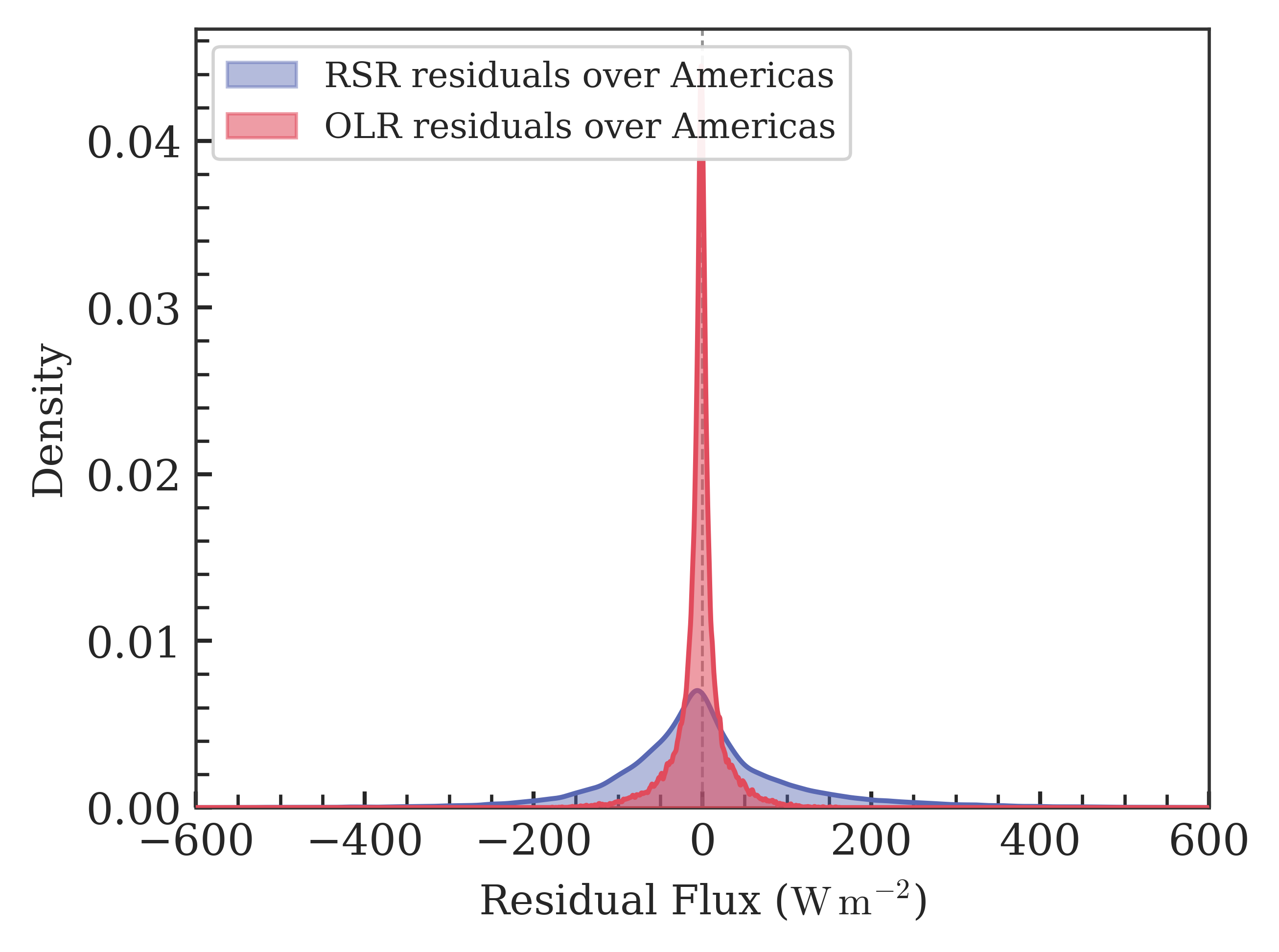}
        \caption{}
        \label{fig:era5_residuals}
    \end{subfigure}
    
    \caption{\textbf{Histograms of OLR and RSR over the Americas}
    \textbf{(a)} Empirical probability density distributions of raw, absolute
    top-of-atmosphere (TOA) Outgoing Longwave Radiation (OLR; red) and Reflected
    Solar Radiation (RSR; blue) fluxes over the Americas domain. The contrast
    between the unconstrained OLR and the heavily skewed, zero-bounded RSR
    motivates modeling the SW flux with a physically scaled treatment term and
    insolation-weighted loss, rather than as an unconstrained linear residual.
    \textbf{(b)} Empirical probability density distributions of the corresponding
    TOA flux residuals after subtracting the meteorological baseline
    ($\text{COIN} - \text{ERA5}$) for OLR (red) and RSR (blue). The narrow,
    symmetric OLR residual distribution lends itself well to a linear Ordinary
    Least Squares regression, the wide and skewed RSR residual
    distribution requires a physically constrained, weighted spline regression
    that adjusts for solar geometry, background contrast, and surface albedo.}
    \label{fig:methodology_combined}
\end{figure}

We model the OLR and RSR effects differently because their physical and statistical properties diverge (Fig.~\ref{fig:methodology_combined}). For OLR, the residuals, calculated by subtracting the ERA5 baseline from COIN observations, are relatively bounded around zero with low variance. This makes it suitable for a simple linear Ordinary Least Squares (OLS) causal regression~\cite{sonabend2026observing}. In contrast, daytime RSR data are more variable, confounded, and skewed~\cite{CHEN201970}. We model RSR using a more sophisticated physics-informed, semi-parametric linear model. We combine flexible cubic B-splines to capture non-linear background weather radiation with a physical treatment term that scales contrail reflection by the available sunlight and the background contrast of the physically underlying scene.

The RSR model also requires more environmental context than the OLR model because solar reflection is physically more complex than thermal emission. While OLR is driven primarily by atmospheric temperatures and clouds, RSR depends heavily on the surface beneath the contrail (surface albedo) and the position of the sun (solar insolation and zenith angle). For example, the same contrail will have a different RSR effect depending on whether it sits over a dark ocean or a bright ice sheet. Without adjusting for these surface and solar variations, background changes in scene brightness would drown out the signal of interest.

We filter for physical daytime observations by requiring  $\theta_0< 80^\circ$ and energy conservation ($\text{RSR}_{\text{ERA5}} \le S$), where $\theta_0$ is the solar zenith angle and $S$ is the ERA5 top-of-atmosphere incident SW flux. Formally, we model expected daytime SW flux $\mathbb{E}[\text{RSR}]$ directly in flux units, combining the advected trace density ($A$) with spline and linear environmental controls:

\begin{equation}\label{rsr_model}
\begin{split}
\mathbb{E}[\text{RSR} \mid A, \mathbf{X}] &= \beta_0 + \beta_1 T_{2\text{m}} + f_{\text{rsr}}(\text{RSR}_{\text{ERA5}}) + f_{\mu}(\mu) + f_{\text{olr}}(\text{OLR}_{\text{ERA5}}) + f_{\text{alb}}(\alpha_{\text{sfc}}) \\
&\quad + \sum_{j=1}^4 \beta_{\text{CP}j} I_{\{\text{CP}=j\}} + \sum_{j=0}^4\left\{ \gamma_j\, I_{\{\text{CP}=j\}} \cdot A\cdot\underbrace{S (1 - \alpha_{\text{sfc}})}_{\text{reflection potential } P}\right\},
\end{split}
\end{equation}

where $\mu = \cos\theta_0$, $\alpha_{\text{sfc}}$ is the MODIS surface albedo described above (with a constant value of 0.08 assigned to water surfaces), and $I_{\{\text{CP}=j\}}$ are indicator variables for the GOES-16 cloud phase, categorized as (0) clear sky, (1) liquid water, (2) supercooled liquid water, (3) mixed phase, or (4) ice. We list each feature explicitly alongside its estimated coefficient for the Americas domain in Table~\ref{tab:rsr_coefficients}. At a high level, eq.~\eqref{rsr_model} uses cubic B-splines to absorb non-linear background variations in ERA5 shortwave flux (64 quantile knots), cosine of the solar zenith angle (20 uniform knots), ERA5 longwave flux (10 uniform knots), and surface albedo (10 uniform knots), while controlling linearly for 2-meter temperature ($T_{2\text{m}}$) and baseline GOES-16 cloud phase, with clear sky ($\text{CP}=0$) as the omitted reference category. All spline bases are constructed without an intercept basis so that the single regression intercept is $\beta_0$, which together with dropping the reference cloud phase leaves a full-rank 118-column design matrix.

\begin{figure}[t]
\centering
\begin{tikzpicture}[
  >={Stealth[length=2.2mm,width=1.7mm]},
  every node/.style={font=\sffamily\small},
  nvar/.style   ={draw=black!70, rounded corners=2pt, align=center,
                  inner sep=3pt, minimum height=9mm, fill=white, line width=0.4pt},
  nadj/.style   ={nvar, fill=black!10},
  ntreat/.style ={nvar, line width=1pt, draw=black},
  nlatent/.style={nvar, dashed, fill=white, draw=black!55},
  nout/.style   ={nvar, line width=1pt, draw=black},
  nderiv/.style ={nvar, draw=black!45, fill=black!4, minimum width=8mm,
                  minimum height=8mm, inner sep=2pt},
  ecausal/.style={->, draw=black!65, line width=0.5pt},
  eweak/.style  ={->, draw=black!55, line width=0.5pt, dashed},
  emod/.style   ={->, draw=black!55, line width=0.5pt, dash pattern=on 1pt off 1pt},
  elab/.style   ={font=\sffamily\scriptsize, text=black!55, inner sep=1.5pt}
]

\node[ntreat]  (a)                      {Advected trace\\[-1pt]density $A$};
\node[nlatent] (c)  [right=18mm of a]   {Contrail cirrus\\[-1pt]{\scriptsize unobserved}};
\node[nout]    (y)  [right=42mm of c]   {Observed RSR\\[-1pt]{\scriptsize COIN}};
\node[nvar]    (t)  [above=16mm of a]   {Time\\[-1pt]{\scriptsize season, local hour}};
\node[nadj]    (g)  [above=16mm of c]   {Solar geometry \& albedo\\[-1pt]{\scriptsize $S,\ \mu,\ \alpha_{\mathrm{sfc}}$}};
\node[nadj]    (w)  [below=16mm of a]   {Meteorology\\[-1pt]{\scriptsize $T_{2\mathrm{m}}$, ERA5 baselines}};
\node[nadj]    (cp) [below=16mm of c]   {Cloud phase\\[-1pt]{\scriptsize ACTP}};

\coordinate (m) at ($(c.east)!0.5!(y.west)$);
\node[nderiv]  (p)  at ($(m)+(0,15mm)$) {$P$};

\draw[ecausal] (t) -- (g);
\draw[ecausal] (t) -- (a);
\draw[ecausal] (t.west) to[out=180,in=180,looseness=1.25] (w.west);
\draw[ecausal] (w) -- (a);
\draw[ecausal] (w) -- (cp);
\draw[ecausal] (a) -- (c);
\draw[ecausal] (c) -- (y);
\draw[ecausal] (g) -- (p);
\draw[ecausal] (g.north) to[out=50,in=100,looseness=1.25] (y.north);
\draw[ecausal] (cp.east) to[out=0,in=-115] (y.south);
\draw[ecausal] (w.south) to[out=-90,in=-90,looseness=0.55] (y.south);

\draw[emod] (p) -- ($(m)+(0,0.7mm)$);
\fill[black!55] (m) circle (1.2pt);

\draw[eweak] (c) -- (cp)
  node[midway, left=3pt, elab] {post-treatment};

\coordinate (lgc) at ($(w.south)!0.5!(y.south)+(0,-26mm)$);
\begin{scope}[shift={(lgc)}]
  \node[nvar,    minimum height=4.5mm, minimum width=7mm, inner sep=1pt] (l1) at (-5.0,0) {};
  \node[elab, text=black!75, anchor=west] at ($(l1.east)+(0.1,0)$) {unadjusted};
  \node[nadj,    minimum height=4.5mm, minimum width=7mm, inner sep=1pt] (l2) at (-2.3,0) {};
  \node[elab, text=black!75, anchor=west] at ($(l2.east)+(0.1,0)$) {adjusted for};
  \node[nlatent, minimum height=4.5mm, minimum width=7mm, inner sep=1pt] (l3) at (0.4,0) {};
  \node[elab, text=black!75, anchor=west] at ($(l3.east)+(0.1,0)$) {unobserved};
  \node[nderiv,  minimum height=4.5mm, minimum width=7mm] (l4) at (3.1,0) {};
  \node[elab, text=black!75, anchor=west] at ($(l4.east)+(0.1,0)$) {deterministic};
\end{scope}

\end{tikzpicture}
\caption{\textbf{Dyrected Acyclic Graph assumed for the RSR causal regression.} Solid arrows denote assumed causal effects. Shaded nodes are adjusted for in Eq.~\eqref{rsr_model}:
solar geometry and surface albedo through splines $f_\mu$ and $f_{\mathrm{alb}}$,
meteorology through $T_{2\mathrm{m}}$ and the ERA5 counterfactual baselines
$f_{\mathrm{rsr}}$ and $f_{\mathrm{olr}}$, and cloud phase through the
indicators. Advected trace density $A$ is an observed proxy for the unobserved
contrails, so the estimand is the flux response per flight kilometer
rather than per contrail. The reflection potential $P = S(1-\alpha_{\mathrm{sfc}})$
is a deterministic function of incident insolation and surface albedo; the
fine-dashed arrow denotes effect modification rather than a causal edge, since
$P$ scales the treatment term and a contrail can only reflect the sunlight that
reaches it. Season and local hour drive both air traffic and solar geometry, and
this backdoor path is blocked by conditioning on $\mu$ and $S$ (see Methods).}
\label{fig:causal_dag_panel}
\end{figure}
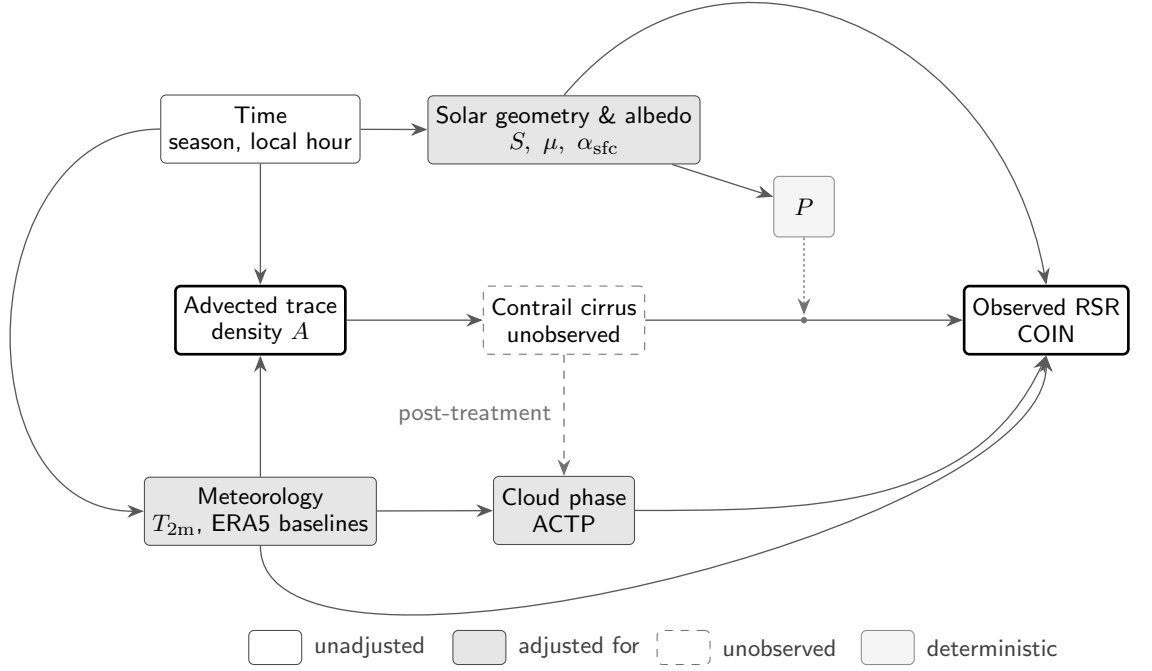

To match our causal graph (Fig.~\ref{fig:causal_dag_panel}), advected trace density is used in the model adjusted through its reflection potential $P \equiv S (1 - \alpha_{\text{sfc}})$, interacting per cloud phase category. The intuition is that a contrail can only reflect the sunlight that reaches it, so its effect scales with incident insolation, and produces the largest marginal reflection when it forms over a dark surface because of high visual contrast, rather than over an already bright one. Allowing $\gamma_j$ to vary by cloud phase lets the sensitivity differ between contrails embedded in ice cloud, where they persist and spread into cirrus, and contrails over liquid or clear scenes. We omit a standalone $A$ term both because it is highly collinear with the sum of the treatment interaction bases, and because it would be physically inadmissible in the SW channel: a treatment effect that persisted as the available sunlight went to zero cannot represent reflection. Note that when advected trace density is zero, the baseline RSR response function to background solar insolation and surface albedo is estimated with the spline bases. All features are standardized to zero mean and unit variance before fitting an $L_2$-regularized (Ridge) regression ($\lambda = 0.05$) to prevent the model from overfitting to complex combinations of weather and exposure.

The contrail SW contrast is only physically detectable where there is both sunlight and advected trace density, so we fit the Ridge regression with weights that scale with the same potential reflection proportional to advected trace density:
\begin{equation*}
w_i \propto 1 + \min\left(2.5, \sqrt{A_iP_i / \overline{AP}}\right), \qquad \bar{w} = 1,
\end{equation*}
where $\overline{AP}$ is the mean of $A_i P_i$ over the daytime pixels, so the weighting is invariant to the overall scale of the reflection potential. This upweights pixels with high reflective forcing potential~\cite{meerkotter1999radiative} and downweights low-insolation pixels that have low signal-to-noise ratios. Pixels with no advected trace density get the floor weight, the cap is enforced when $A_i P_i > 2.5^2\,\overline{AP}$, so the resulting maximum-weight to floor ratio is $3.5{:}1$. This follows standard weight-trimming practice in weighted regression~\cite{kish1965survey, potter1990study} to help control the variance, so that a small cluster of high-insolation, high-advected trace density pixels cannot dominate the regression fit loss. Normalizing the weights to unit mean leaves the effective ratio between the loss and the $L_2$ penalty unchanged. These weights affect only the fit; the ATE remains the unweighted population effect over the empirical distribution of pixels.

Finally, the daytime Average Treatment Effect (ATE; in $\text{GJ/km}$) is estimated by comparing counterfactual predictions between observed advected trace density ($A$) and a perturbed state ($A + \bar{A}$), where $\bar{A}$ is the mean advected trace density across the daytime dataset. Averaging this prediction difference over pixels and dividing by $\bar{A}$ gives the marginal flux response per unit of advected trace density, which is then converted from $\text{W}\,\text{m}^{-2}$ per unit of advected trace density into energy forcing per flight kilometer (for further details refer to \cite{sonabend2026observing}). Since nighttime contrails produce no solar reflection, this daytime effect is scaled by the daytime fraction of total flight kilometers across the spatio-temporal domain, to estimate mean SW forcing comparable to mean LW forcing.

Treatment regression coefficients are expressed in reflection potential units, so each $\gamma_j$ is the marginal reflected flux per unit of trace density, per unit of available sunlight, over a perfectly dark (albedo=0) background surface; a positive value therefore means cooling (larger amount of sunlight reflected). Note however, in all locations in this work when reporting the final oRF values (LW, SW and Net), we adopt the standard earth-system convention of "positive sign means warming to the earth system".

The aggregate daytime effect is dominated by contrails over supercooled liquid ($\hat\gamma_2 = 3.85 \pm 0.19$) and mixed-phase ($\hat\gamma_3 = 1.57 \pm 0.18$) scenes, with a smaller but tightly constrained clear-sky contribution ($\hat\gamma_0 = 0.81 \pm 0.07$). In ice cloud scenes, where contrails are optically embedded in (or above) pre-existing cirrus, the shortwave sensitivity is statistically indistinguishable from zero ($\hat\gamma_4 = -0.06 \pm 0.20$); this is expected since the marginal reflectance of an added ice layer is small when the background scene is already bright and ice-covered. Over liquid water cloud the estimate is small and negative ($\hat\gamma_1 = -1.26 \pm 0.18$), consistent with the reduced contrast of an ice layer overlying bright low cloud~\cite{meerkotter1999radiative}. Together these coefficients determine how the contrail shortwave effect varies across solar angles, surface brightness and background scene conditions.

\begin{table}[htbp]
\centering
\caption{Estimated coefficients for the RSR causal regression model~\eqref{rsr_model} over the Americas domain, from a 30-million pixel sample of 363 days ($N = 6{,}424{,}372$ daytime training pixels, $L_2$ penalty $\lambda = 0.05$). Coefficients are reported on standardized features, so each entry is the change in reflected flux ($\text{W}\,\text{m}^{-2}$) per standard deviation of the covariate; parentheses give the standard deviation over 1,000 day-level block bootstrap replicates. The 108 cubic B-spline basis coefficients absorbing the background environment ($\text{RSR}_{\text{ERA5}}$, $\mu$, $\text{OLR}_{\text{ERA5}}$, $\alpha_{\text{sfc}}$) are omitted for brevity.}
\label{tab:rsr_coefficients}
\begin{tabular}{lcr}
\toprule
\textbf{Covariate} & \textbf{Model Term} & \textbf{Estimate (SD)} \\
\midrule
\multicolumn{3}{l}{\textit{Baseline Scene}} \\
Baseline offset & $\beta_0$ & $206.782$ $(0.783)$ \\
Background splines: $\text{RSR}_{\text{ERA5}}$ (65), $\mu$ (21), & \multirow{2}{*}{$f_{\text{rsr}}, f_{\mu}, f_{\text{olr}}, f_{\text{alb}}$} & \multirow{2}{*}{\textit{omitted}} \\
\quad $\text{OLR}_{\text{ERA5}}$ (11), $\alpha_{\text{sfc}}$ (11) & & \\
$T_{2\text{m}}$ (2-meter temperature) & $\beta_1$ & $-5.000$ $(0.301)$ \\
\midrule
\multicolumn{3}{l}{\textit{Cloud Phase Intercepts (reference: CP $=0$, clear sky)}} \\
$I_{\{\text{CP}=1\}}$ (liquid water) & $\beta_{\text{CP}1}$ & $31.200$ $(0.282)$ \\
$I_{\{\text{CP}=2\}}$ (supercooled liquid) & $\beta_{\text{CP}2}$ & $27.275$ $(0.310)$ \\
$I_{\{\text{CP}=3\}}$ (mixed phase) & $\beta_{\text{CP}3}$ & $21.963$ $(0.249)$ \\
$I_{\{\text{CP}=4\}}$ (ice) & $\beta_{\text{CP}4}$ & $48.026$ $(0.275)$ \\
\midrule
\multicolumn{3}{l}{\textit{Treatment: $A \cdot P$, with reflection potential $P = S (1 - \alpha_{\text{sfc}})$}} \\
$A P \cdot I_{\{\text{CP}=0\}}$ (clear sky) & $\gamma_0$ & $0.814$ $(0.067)$ \\
$A P \cdot I_{\{\text{CP}=1\}}$ (liquid water) & $\gamma_1$ & $-1.260$ $(0.177)$ \\
$A P \cdot I_{\{\text{CP}=2\}}$ (supercooled liquid) & $\gamma_2$ & $3.853$ $(0.189)$ \\
$A P \cdot I_{\{\text{CP}=3\}}$ (mixed phase) & $\gamma_3$ & $1.565$ $(0.178)$ \\
$A P \cdot I_{\{\text{CP}=4\}}$ (ice) & $\gamma_4$ & $-0.060$ $(0.199)$ \\
\bottomrule
\end{tabular}
\end{table}

\subsection*{Controlling for Correlation Using Block-Bootstrap}

Satellite pixel data on any given day are influenced by the same large-scale synoptic weather systems, so the background baseline and contrail radiative effect are naturally correlated. Treating each pixel as an independent observation would ignore this shared weather context and artificially narrow our uncertainty bounds. To account for this correlation, we use a day-level block bootstrap procedure~\cite{sonabend2026observing}.

For each bootstrap set, we randomly sample entire days with replacement from our complete dataset, then sample pixels within the sampled days. We run both the OLR and RSR regressions using the exact same set of pixels from these sampled days. This paired approach ensures that both treatment effects are calculated over identical scenes, allowing us to combine them and compute a physically consistent net treatment effect per sample (\orfnet = \orfsw + \orflw).

We repeat this entire sampling and estimation process 1,000 times to construct the final empirical bootstrap distributions. The 95\% confidence intervals for the LW, SW, and net forcing estimates are then derived directly from the 2.5th and 97.5th percentiles of these 1,000 runs.

\subsection*{Automated Feature Selection and Synthetic Validation}

To systematically search the space of models that fit within our DAG and select the spline configuration and Ridge regularization strength of the SW model, we used the automated code generation system introduced by Aygün et al.~\cite{aygun2025automated}. This framework couples Large Language Models (LLMs), specifically utilizing Gemini 3 Flash as the generative backbone for iterative, syntax-valid code synthesis with a Monte Carlo Tree Search (MCTS) via the Empirical Research Assistance system to sample and optimize target performance surfaces. Guided by the Predictor Upper Confidence Bound (PUCB) heuristic~\cite{silver2016mastering}, the pipeline functions as an Automated Machine Learning system~\cite{hutter2019automated} embedded with domain-specific physical constraints. 

We supplied the optimization loop with physical constraints. To prevent collider bias, the feature pool provided to the MCTS loop was strictly limited to meteorological conditions and treatment confounder variables. From this safe feature space, the MCTS loop systematically optimizes the selections from a list of provided confounders and interaction terms, as well as a regularization grid and possible feature transformations (such as sine and cosine solar zenith angle, and polynomials).

This automated feature engineering is executed within a synthetic validation environment comprised of several synthetic datasets (delineated in the main text, \hyperref[isolating_signals]{Isolating Aviation Signals from Natural Variability}) that served as a controlled evaluation sandbox. Because this synthetic environment explicitly conserves the joint distribution of meteorological confounders found in the raw observational data, conducting the structural search within this sandbox prior to real-world application serves as a guardrail against overfitting. This ensures that our causal model is in fact working on the simulated datasets before running it on real observation data where the treatment effect we estimate was previously unknown.

\backmatter

\bmhead{Acknowledgements}

The authors would like to gratefully acknowledge Tharun Sankar and Aaron Sarna for their software contributions that aided this work, John Platt for his insightful guidance and support, Sebastian Eastham for helpful early discussions, Dinesh Sanekommu for helpful comments on the manuscript, and Erica Brand and Rachel Soh for their contributions in data acquisition.

\section*{Declarations}

\textit{Conflict of interest}. Authors are employees of Google Inc. Google is a technology company that sells computing and machine learning services as part of its business.

\textit{Data availability.} ERA5 data are available from the Copernicus Climate Change Service Climate Data Store (CDS) https://doi.org/10.24381/cds.bd0915c6~\cite{hersbach2020era5}. COIN data can be found at https://console.cloud.google.com/storage/browser/upwelling\_irradiance/ceres\_goes/ (last access: 30 June 2025). GOES-16 products can be found at https://console.cloud.google.com/marketplace/product/noaa-public/goes (last access: 30 June 2025). 





\textit{Author contribution.} ASW: conceptualization, software, visualization, writing – original draft, review and editing. SG: conceptualization, software, writing – review and editing. NG: software, writing – review and editing, supervision. JN: software. CVA: conceptualization, writing – review and editing, supervision. KM: conceptualization, software, visualization, writing –review and editing.









\bibliography{sn-bibliography}

\end{document}